\documentclass[a4paper]{article}

\usepackage{icrc2013}

\title{Search for radio counterpart of the Fermi/VERITAS PWN candidate in the SNR CTA 1}

\shorttitle{Search for radio counterpart of the PWN candidate in the SNR CTA 1}

\authors{
E. Giacani$^{1,2}$,
A.C. Rovero$^{1}$,
A. Cillis$^{1}$
A. Pichel$^{1}$
G. Dubner$^{1}$,
}

\afiliations{
$^1$ Instituto de Astronom\'ia y F\'isica del Espacio (CONICET-UBA), Buenos Aires, Argentina \\
$^2$ FADU, University of Buenos Aires, Buenos Aires, Argentina\\
}

\email{egiacani@iafe.uba.ar; rovero@iafe.uba.ar}

\abstract{
We present new high angular resolution and high sensitivity radio observations toward the neutron star RX J0007.0+7303, carried out with the Karl G. Jansky Very Large Array at 1.5 GHz. This source powers a pulsar wind nebula (PWN) only detected in the X-ray and gamma-ray domains. The new high quality radio observations do not show any evidence of a source, either point-like or extended, that could be interpreted as the radio counterpart of the high energy PWN, down to a noise level of 15 $\mu$Jy/beam.
}


\begin{document}
\maketitle

%Begin a section.
\section{Introduction}

Among the rich population of gamma-ray galactic sources, pulsar
wind nebulae (PWNe) constitute the largest set of galactic TeV emitters with more than 30 sources of this class detected. In the past 30 years, observations in radio, X-ray, and gamma-ray bands with
growing sensitivity have notably increased the number of pulsar-supernova remnants (SNRs) associations.
{\it Fermi/LAT} observations have remarkably contributed to this fact after the discovery
of several gamma-ray pulsars not detected previously at other wavelengths.
One of these detections is the pulsating emission associated with the point source
RX J0007.0+7303 [(J2000.0) $00^h07^m01^s.56, +73^{\circ}03^{\prime}08^{\prime\prime}.1$]
in the SNR CTA 1 (G119.5+10.2), discovered by {\it Fermi/LAT} during its initial
operation and later confirmed as a pulsar in X-rays.

CTA 1 was first proposed as a SNR by Harris \& Roberts \cite{bib:harris}. In the radio band,
this SNR is approximately circular in shape $\sim \! 90^{\prime}$ in diameter,
surrounded by a bright thin shell along the eastern and southern border.

In the X-ray domain, the first detection of the SNR was based on {\it ROSAT} observations,
where centrally-peaked emission was detected \cite{bib:seward}.
The authors also reported the presence of a faint compact source, RX J0007.0+7303, located
within the brightest central region. The X-ray spectrum requires both a thermal
component, that dominates the outer regions of the remnant, and a power-law component,
that dominates the central emission. Subsequently, the non-thermal nature of the central emission
was confirmed using {\it ASCA} observations revealing a syncrotron nebula of 18 arcmin in diameter. These data showed also the existence of extended emission, up to  $\sim$ 40 arcmin in radius, likely of synchrotron origin \cite{bib:slane1, bib:slane2, bib:halpern}.
These early detections were indicative of the presence of a synchrotron
nebula driven by an active neutron star, for which RX J0007.0+7303 represented a viable
pulsar candidate.

Further studies performed with the {\it XMM-Newton} and {\bf{\it ASCA}} satellites towards RX J0007.0+7303 resolved the X-ray emission
into a point-like source and a diffuse nebula, \cite{bib:slane2}.
The spectrum is well fitted with two components, a blackbody
plus a power-law, consistent with the picture of a young pulsar surrounded by a nebula. 

The field
has been also investigated using {\it Chandra} X-ray observatory \cite{bib:halpern}, whose
higher resolution image revealed the point source, RX J0007.0+7303 embedded in a compact
nebula $3^{\prime\prime}$ in radius and a jet-like extension. The most plausible
interpretation of these data is that a young and energetic pulsar is powering a
synchrotron PWN.

At high energy (HE) gamma-rays, the EGRET source 3EG J0010+7309 (in spatial coincidence with RX
J0007.0+7303) was discussed as a potential candidate for a radio-quiet gamma-ray pulsar \cite{bib:mattox}.
Other authors also proposed that this source showed the behavior of a pulsar \cite{bib:brazier}, but
a search for gamma-ray pulsation using EGRET data failed to reveal this characteristic \cite{bib:ziegler}.

Data acquired during the commissioning phase of {\it Fermi/LAT} (30 June to 30 July 2008)
and in the initial days of routine operation (5 to 20 August 2008), revealed a 315.87 ms
radio-quiet pulsar in CTA 1 \cite{bib:abdo1}. The new
pulsar in CTA 1 exhibits all the characteristics of a young HE pulsar. The spin-down
power of $4.5 \times 10^{35}$ erg s$^{-1}$ is sufficient to supply the PWN with magnetic
fields and energetic electrons at the required rate of $10^{35}$ to $10^{36}$
erg $s^{-1}$, and the pulsar age $\tau_{c}$ $=$ 14 000 yr \cite{bib:abdo1} is consistent with the age of the host SNR CTA 1 estimated to be in the range between 5000 and 15000 yr \cite{bib:pineault, bib:slane1, bib:slane2}. 
According to {\it Fermi/LAT} observations the pulsar surface magnetic field is $\sim$ 1.1$\times 10^{13}$ G, the second highest among known gamma-ray pulsars. The first one is PSR J1509-58 with an inferred field of 1.54$\times 10^{13}$ G that shows emission up to $\sim$ 300 MeV, while emission from CTA 1 pulsar is present to at least 5 GeV \cite{bib:abdo1}.

Following this discovery, {\it XMM-Newton} observations were carried out
searching for pulsation in X-rays, finally detected in 2010 \cite{bib:lin, bib:caraveo}.
The detected pulsed period is consistent with the gamma-ray periodicity.

A candidate for an off-pulse emission from RX J0007.0+7303 was noted using 16 months
of {\it Fermi/LAT} data \cite{bib:ackermann}, but it was below the detection threshold.
With more data accumulated, the detection of an extended source in the off-pulse emission at
$\sim \! 6 \sigma$ level was reported using 2 years of {\it Fermi/LAT} data \cite{bib:abdo2}.
This extended source is a potential PWN, although the possibility that it originates inside the
magnetosphere can not be ruled out.

At VHE gamma-rays the VERITAS Collaboration also observed the source during 25.5 hours of quality-selected
observations in the period October 2010 - January 2011, and 15.5 additional hours in the period
September-November 2011. They reported a post-trials significance of 6.5 $\sigma$ detection of an extended
source with a centroid near the {\it Fermi/LAT} gamma-ray pulsar and its X-ray PWN \cite{bib:aliu}.
The centroid of their detection zone is 5 arcmin from the position of RX J0007.0+7303, which is coincident
within uncertainties. The resulting 1$\sigma$ angular extent is $0.30^{\circ} \pm 0.03^{\circ}$ along the semi-mayor
axis and $0.24^{\circ} \pm 0.03^{\circ}$ along the semi-minor axis. The properties of the detected extended source
fit nicely with the known TeV/X-ray PWN population.

No radio counterpart to RX J0007.0+7303 was identified in a list of compact sources
reported by Pineault et al. \cite{bib:pineault}. Deep searches of radio emission, either pulsating
or not, performed with the Very Large Array in its A configuration at 1.4 GHz,
and with the NRAO Green Bank Telescope at 820 MHz, have yielded negative results
\cite{bib:halpern}. If RX J0007.0+7303 is indeed a radio pulsar, its radio
luminosity is an order of magnitude below the faintest radio pulsars known so far,
or its radio beam does not intersect the Earth.

The kinematic distance to the system of 1.4 $\pm$ 0.3 kpc has been derived based on the asssociation of an HI shell with the SNR \cite{bib:pineault}.

In this paper we present new high-resolution and high-sensitivity radio observations
conducted with the Karl G. Jansky Very Large Array (JVLA) instrument in a broadband around 1.5 GHz to search
for a radio counterpart to the PWN driven by RX J0007.0+7303, observed at high and very high energy bands.

%-----------------------------------------------------------------------
\section {Radio continuum observations and results}

The radio continuum observations toward the source RX J0007.0+7303 were performed, under the shared
risk commissioning phase, with the JVLA\footnote{The
Very Large Array of the National Radio Astronomy Observatory is a facility of the National
Science Foundation operated under cooperative agreement by Associated Universities, Inc.} in its B
configuration, on 27 July 2012 for a total of 60 minutes on-source integration time. We used the
full 1 GHz bandwidth covering the frequency range from 1 to 2 GHz centred at 1.5 GHz. The data were
recorded in 16 spectral windows (SPWs) with a bandwidth of 128 MHz each, spread into 64 channels.
Ten channels at the edges of the SPWs were flagged due to the roll-off of the digital filters in the
signal chain. To reduce the size of the data, after initial data flagging the data were averaged to 5 s
resolution time.

Data processing was carried out using the CASA and MIRIAD software packages, following standard procedures.
The source 3C48 was used for primary flux density and bandpass calibration, while phases were calibrated with J0019+7327.
The image was deconvolved using natural weighting of the visibility data, which gives an increased sensitivity for extended
structures, although it results in a larger synthesized beam. This was constructed with the task MAXEN in MIRIAD,
which performs a maximum entropy deconvolution algorithm on a cube.
The resulting image, shown in Figure 1, has a synthesized beam of 6.5$^{\prime\prime} \times 4.4^{\prime\prime}$,
and an rms noise of about 15 $\mu$Jy/beam. This new image toward RX J0007.0+7303 improves the level of noise in a
factor two with respect to the previously published image at the same frequency.

\begin{figure}
\includegraphics[width=0.5\textwidth]{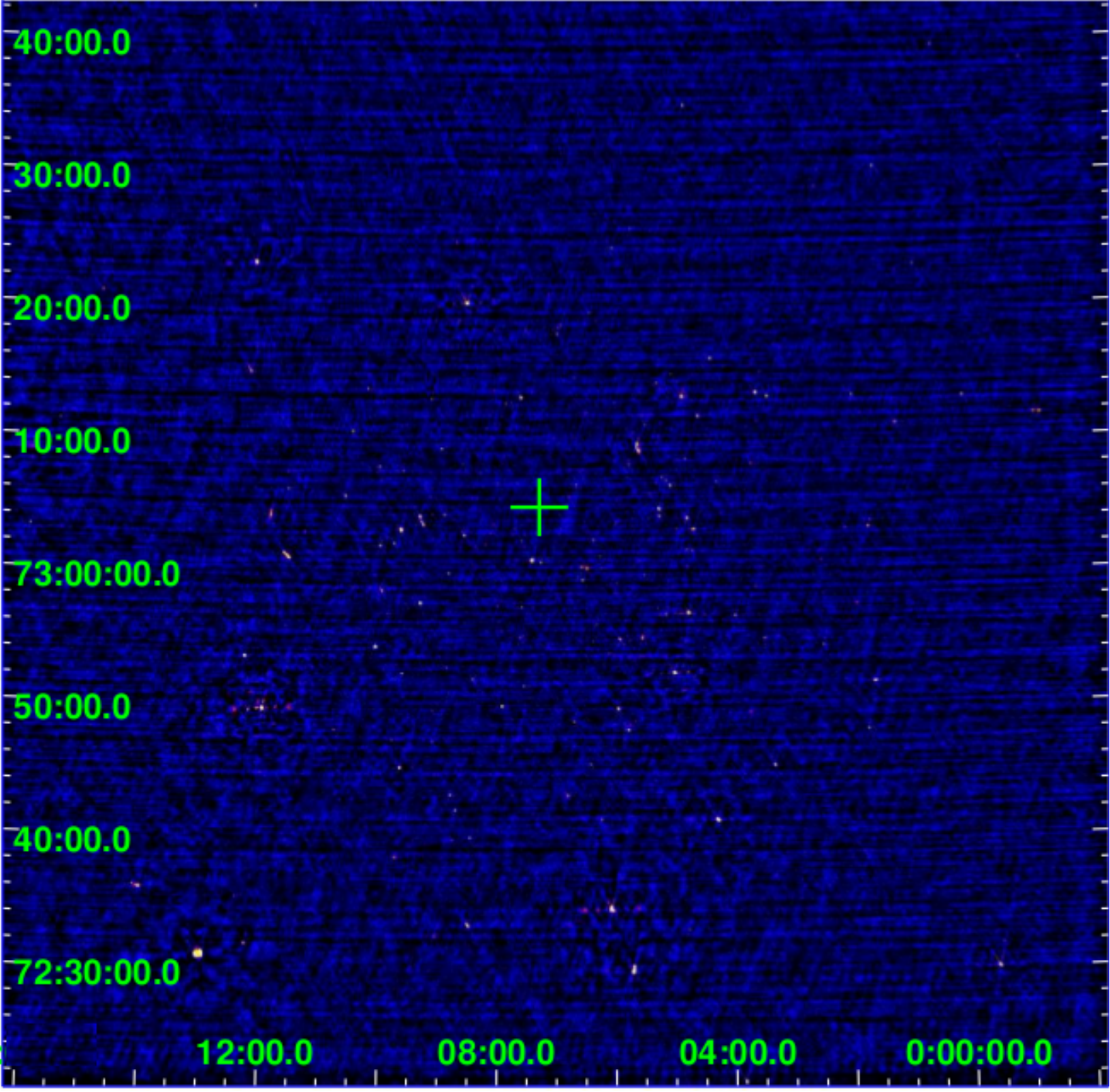}
\caption{JVLA image of the observed field toward RX J0007.0+7303 at 1.5 GHz. The image has a synthesized beam  of
6$^{\prime\prime}.5 \times 4^{\prime\prime}$.4, and an rms noise of 15 $\mu$Jy/beam. The plus sign indicates the position of the neutron star.}
\label{fig:radio}
\end{figure}

\begin{figure}
\includegraphics[width=0.5\textwidth]{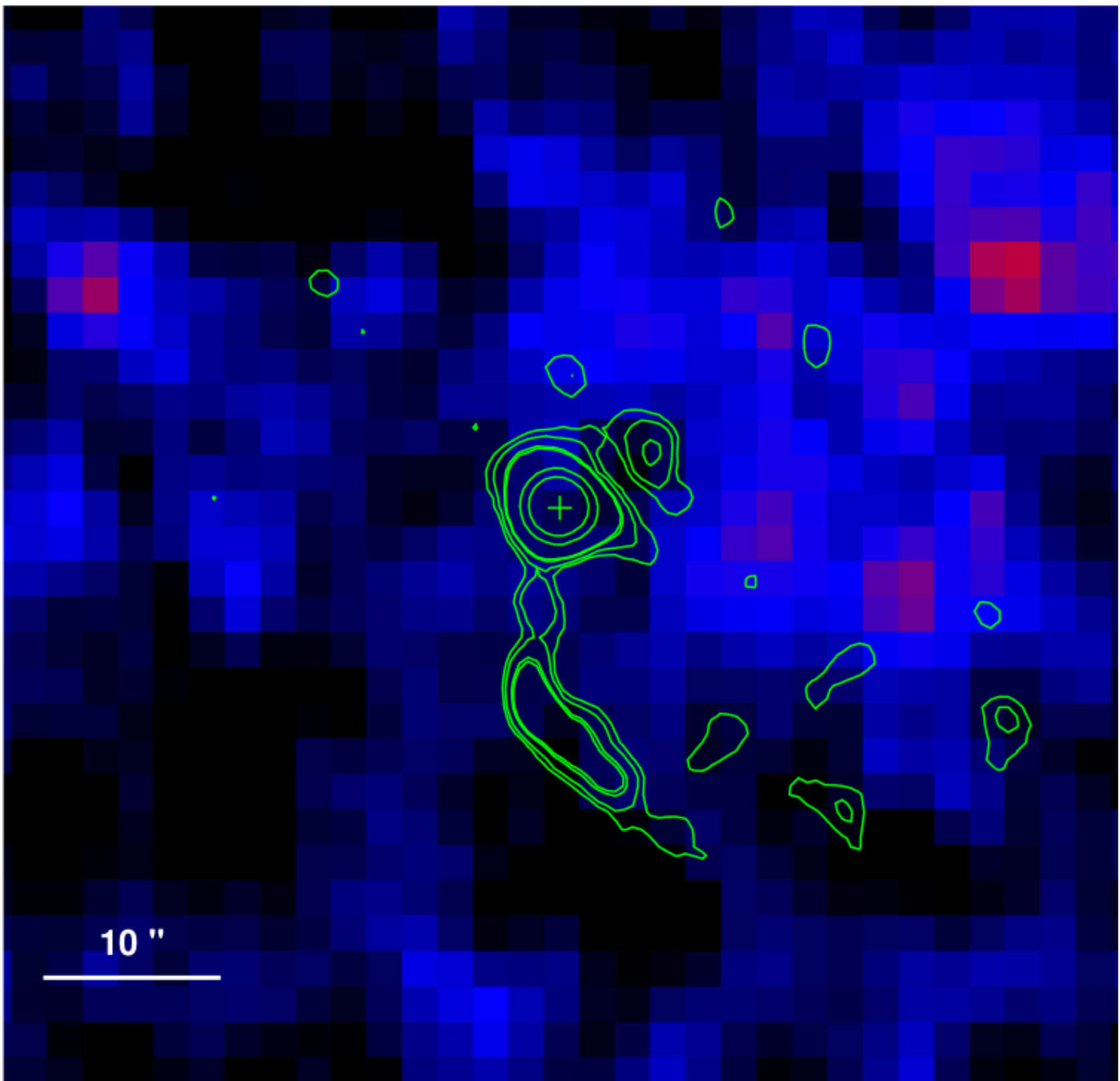}
\caption{An enlargement of the JVLA radio image around RX J0007.0+7303 with some contours of the {\it Chandra} image in the 0.5-8 keV band. The plus sign indicates the position of the neutron star.}
\label{fig:rx}
\end{figure}

\begin{figure*}[t!]
\centering
\includegraphics[width=0.7\textwidth]{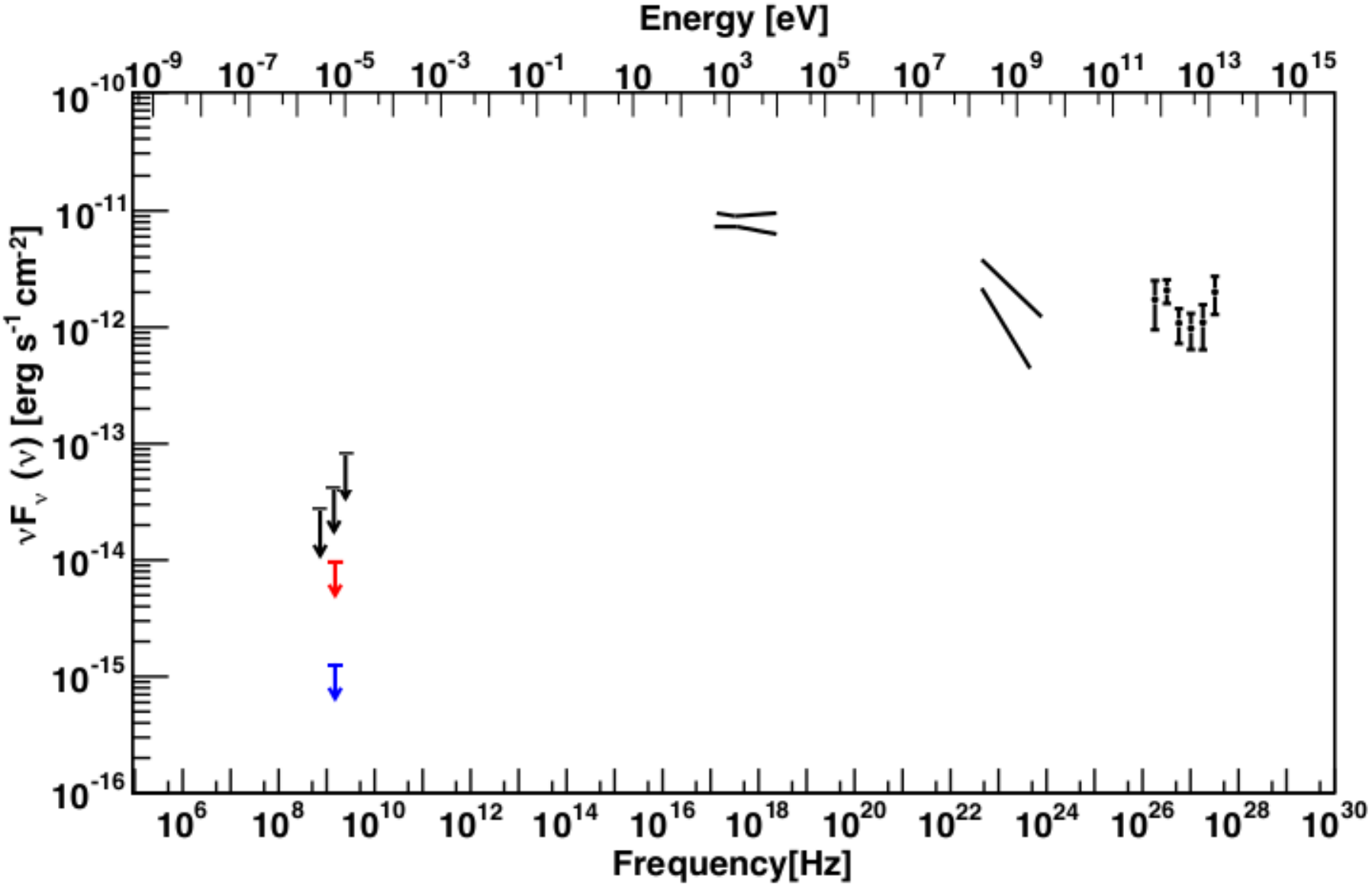}
\caption{Spectral energy distribution. Very high energy observations are from \cite{bib:aliu}, high energy observations from \cite{bib:abdo2}; X-Ray observations from \cite{bib:slane1, bib:slane2}. The radio upper limits as considered by VERITAS \cite{bib:aliu} are plotted in black. The results from our work are plotted in red (blue) assuming a size of 20 (9) arcmin in radius for the PWN.}
\label{fig:sed}
\end{figure*}

Figure 2 displays an enlargement of the central field toward RX J0007.0+7303 with some contours of the PWN as seen in the
{\it Chandra} image. The lack of both extended and point-like emissions is evident in the figure, preventing us to make any association
with RX J0007.0+7303 and its nebular emission detected in the X-ray and gamma-ray bands. Therefore, we can only determine
upper limits to the radio flux density, for which an estimation of the size of the nebula is needed. 

If we assume that the size of the PWN in the radio band is at least the
same as that of the X-ray synchrotron nebula as observed with the {\it ASCA} satellite (18 arcmin in diameter, \cite{bib:halpern}), we obtain a flux of $\sim$0.09 Jy
(equivalent to $\nu S_{\nu} = 1.3 \times 10^{-15}$ erg cm$^{-2}$ s$^{-1}$). This estimate  represents the lowest value for the upper limit of the radio flux density. In the case that the "putative" radio PWN has a same size of that observed in the VHE gamma-ray domain \cite{bib:aliu} (a circle of 40 arcmin in diameter), the upper limit for the radio flux density is $\sim$0.7 Jy ($1.0\times 10^{-14}$ erg cm$^{-2}$ s$^{-1}$), approximately a factor five lower than the radio flux density considered by VERITAS.
In Figure \ref{fig:sed} we present the spectral energy distribution. The observational data was adquired as follows: TeV data was obtained from \cite{bib:aliu}; {\it Fermi/LAT} data, from \cite{bib:abdo2}; X-ray observations are from \cite{bib:slane2}; the radio upper limits obtained from \cite{bib:aliu} are also plotted for comparison; the results of our work are in red (UL considering a size for the nebula of 20 arcmin in radius, and blue, UL considering a size for the nebula of 9 arcmin of radius.

\section {Summary and discussion}

We report on the highest sensitivity radio survey of a large region in direction to the neutron star RX J0007.0+7303. The observations were carried out in a broad-band centred at 1.5 GHz using  the Jansky VLA in its B configuration. Despite the good quality of our new radio image (noise level $\sim $15 $\mu$Jy/beam) neither a point-like radio source nor an extended emission were found as a counterpart of the pulsar and its pulsar nebula detected at high and very high energies. Based on this non-detection, we have set upper limits on the radio emission considering two possible sizes for the nebula. 
We considered that the radio nebula, if it exits, should have a size between that estimated for the X-ray synchrotron nebula and that of the VHE PWN. Under these assumptions the upper limit for the flux density obtained from our radio measurements ranges from 1.3$\times$10$^{-15}$ to 1.0$\times$10$^{-14}$ erg~cm$^{-2}$~s$^{-1}$.

Previous radio observations of the SNR CTA 1 \cite{bib:pineault} were used to predict the gamma emission of the source using a leptonic model \cite{bib:zhang}. They obtained an upper limit for the radio flux density assuming that the entire emission from the SNR was associated to the PWN. This assumption led the authors to over predict the gamma emission seen by VERITAS \cite{bib:aliu}.
The VERITAS  Collaboration, instead, estimated the radio flux density at 1.4 GHz using for the nebula a radius of 20 arcmin. They also extrapolated this value to lower and higher frequencies assuming a spectral index  $\alpha=0.3$ for lower frequencies and 0 for higher frequencies (S $\propto \nu^{-\alpha}$). In this case, their model fits well  with VERITAS observations.  

We note that our estimation of the radio flux upper limits are even lower than the ones considered by VERITAS \cite{bib:aliu} by a factor of 5, if we assume the same extension from the radio emission, and by more than one order of magnitude if we assume that the size of the radio emission is the same that the synchrotron nebula seen by {\it ASCA}. It is important to mention that our observations do not include either diffuse background emission nor a possible contribution by the shell. Moreover, we have subtracted the contribution of bright point-like sources.

Even if we have derived two values for the upper limit of the radio flux density, the remarkable conclusion is that the radio nebula was not detected in the band 1-2 GHz with an rms noise of about 15 $\mu$Jy/beam. Although in this particular case, the high magnetic field of RX J0007.0+7303, may be inhibiting the production of synchrotron radiation at longer wavelengths, there are other cases where the search for radio PWNe yielded negative results, such as for example in the SNRs Kes 79 and G338.3-0.0 \cite{bib:giacani, bib:castelletti}, around the pulsar PSR J1826-1334 \cite{bib:gaensler}, etc. Even though they are objects with different characteristics, the non-detection of synchrotron radiation in the radio band with the current instruments would indicate the necessity to investigate a physical mechanism 
that produces only a PWN observed at high and very high energy photons.

\vspace*{0.5cm}
\footnotesize{{\bf Acknowledgment:}{
This research was partially funded by Argentina Grants awarded by ANPCYT:PICT0571/11; CONICET 0736/12 and University of Buenos Aires (UBACYT20020100100011). The following authors are members of ``Carrera del Investigador Cient\'ifico'' of CONICET, Argentina: E.G., A.C.R., A.C. and G.D.
}

\end{document}